\RequirePackage{fix-cm}

\documentclass[letterpaper, 10 pt, conference]{ieeeconf}

\IEEEoverridecommandlockouts

\def\BibTeX{{\rm B\kern-.05em{\sc i\kern-.025em b}\kern-.08em
    T\kern-.1667em\lower.7ex\hbox{E}\kern-.125emX}}

\usepackage[T1]{fontenc}

\usepackage{etoolbox}

\makeatletter
\def\input@path{{settings}} 
\usepackage{settings/isasmathmacros} 
\makeatletter

\usepackage{nicefrac}
\usepackage{array}
\usepackage{booktabs}
\usepackage[linesnumbered,ruled,vlined]{algorithm2e}
\AtBeginEnvironment{algorithm}{\DontPrintSemicolon}
\SetAlCapNameFnt{\small}
\SetAlCapFnt{\small}

\usepackage[x11names, svgnames, rgb]{xcolor} 
\usepackage{graphics} 
\usepackage[]{graphicx}  
\graphicspath{ {./images/**} } 

\usepackage[font=normalsize,skip=1pt, font=small]{caption}
\usepackage[skip=1.5pt, font=small]{subcaption} 

\usepackage{balance} 

\usepackage{stfloats} 

\usepackage{url} 

\usepackage{nohyperref}

\usepackage{cite} 

\usepackage{cleveref}

\crefname{figure}{Fig.}{Figs.} 
\crefname{section}{Sec.}{Secs.} 
\crefname{table}{Tab.}{Tabs.} 
\crefname{algorithm}{Alg.}{Algs.} 
\crefname{equation}{}{} 
\crefname{appendix}{Appx.}{Appx.}

\crefname{assumption}{Assump.}{Assump.} 

\crefname{myAssumptionCtr}{Assump.}{Assump.} 

\crefname{theorem}{Theorem}{Theorems} 
\crefname{myTheoremCtr}{Theorem}{Theorems} 

\crefname{example}{Example}{Examples} 
\crefname{myExampleCtr}{Example}{Examples} 

\expandafter\def\csname ver@etex.sty\endcsname{3000/12/31}

\usepackage{autonum}

\usepackage{siunitx}
\usepackage{microtype}
\usepackage[all]{nowidow}
\usepackage[subtle,leading,leadingfraction=0.97,mathspacing=normal,wordspacing=normal ]{savetrees}

\usepackage{bbold} 

\newcommand{\ie}{i.e.}

\newcommand{\eg}{e.g.}
\newcommand{\Eg}{E.g.}

\definecolor{myblue}{named}{RoyalBlue}

\definecolor{mygreen}{named}{LimeGreen}

\newcommand{\pdf}{\ensuremath{f}} 

\newcommand{\horizon}{\ensuremath{N}_{\mathrm{H}}} 

\newcommand{\numCEMIter}{\ensuremath{j}_\mathrm{max}} 
\newcommand{\numCEMSamples}{\ensuremath{N}} 

\newcommand{\optVarCemScalar}{\ensuremath{\xi}} 
\newcommand{\vVarCem}{\ensuremath{\vec{\optVarCemScalar}}} 
\newcommand{\dimVarCem}{\ensuremath{d_{{\optVarCemScalar}}}} 
\newcommand{\evVarCem}{\ensuremath{\mean{\vec{\optVarCemScalar}}}} 
\newcommand{\paramCEM}{\ensuremath{\vtheta}} 

\newcommand{\lcdVarScalar}{\ensuremath{\xi}} 
\newcommand{\vVarLcd}{\ensuremath{\vec{\lcdVarScalar}}} 

\newcommand{\bufferSize}{\ensuremath{N_{\mathrm{Buffer}}}} 

\usepackage[]{acro}

\acsetup{single=1} 
\acsetup{single-style=long} 

\DeclareAcronym{AL}         {short={AL},	    	long={active learning}}
\DeclareAcronym{ANEES}      {short={ANEES},	    	long={averaged normalized estimation error squared}}
\DeclareAcronym{ANLL}       {short={ANLL},	    	long={average negative log-likelihood}}

\DeclareAcronym{BNN}		{short={BNN},	    	long={Bayesian neural network}}

\DeclareAcronym{CDF}		{short={CDF},	    	long={cumulative density function}}
\DeclareAcronym{CEM}		{short={CEM},	    	long={cross-entropy method}}
\DeclareAcronym{CMA-ES}	    {short={CMA-ES},	    long={covariance matrix adaptation evolution strategy}, long-plural-form={covariance matrix adaptation evolution strategies}}
\DeclareAcronym{CvM}		{short={CvM},	    	long={Cram\'er--von Mises}}

\DeclareAcronym{DDP}		{short={DDP},	    	long={differential dynamic programming}}
\DeclareAcronym{DM}	        {short={DM},	    	long={Dirac mixture}}
\DeclareAcronym{DMD}		{short={DMD},	    	long={Dirac mixture density}}
\DeclareAcronym{dsCEM}      {short={dsCEM},         long={deterministic sampling CEM}}

\DeclareAcronym{ECE}	    	{short={ECE},	    	long={expected calibration error}}
\DeclareAcronym{EKF}		{short={EKF},	    	long={Extended Kalman Filter}}
\DeclareAcronym{EM}	    	{short={EM},	    	long={expectation--maximization}}
\DeclareAcronym{ENCE}	    	{short={ENCE},	    	long={expected normalized calibration error}}
\DeclareAcronym{EP}	    	{short={EP},	    	long={expectation propagation}}

\DeclareAcronym{GM}	        {short={GM},	    	long={Gaussian mixture}}
\DeclareAcronym{GEVR}	    {short={GEVR},	    	long={generalized error variance ratio}}
\DeclareAcronym{GUCE}		{short={GUCE},	    	long={generalized \acs{UCE}}}

\DeclareAcronym{iLQR}		{short={iLQR},	    	long={iterative linear quadratic regulator}}
\DeclareAcronym{kdtree}		{short={\mbox{$k$-d~tree}},	    	long={$k$-dimensional tree}}
\DeclareAcronym{KBNN}		{short={KBNN},	    	long={Kalman Bayesian Neural Networks}}
\DeclareAcronym{KKT}        {short={KKT},	    	long={Karush--Kuhn--Tucker}}
\DeclareAcronym{KL}         {short={KL},            long={Kullback--Leibler}}
\DeclareAcronym{KS}		    {short={KS},	    	long={Kolmogorov--Smirnov}}

\DeclareAcronym{LCD}		{short={LCD},	    	long={localized cumulative distribution}}
\DeclareAcronym{LRKF}	    {short={LRKF},	    	long={Linear Regression Kalman filter}}

\DeclareAcronym{MAP}	    {short={MAP},	    	long={maximum a posteriori}}
\DeclareAcronym{MC}	        {short={MC},	    	long={Monte Carlo}}
\DeclareAcronym{MCMC}	    {short={MCMC},	    	long={Markov chain Monte Carlo}}
\DeclareAcronym{mCvM}		{short={mCvM},	    	long={modified Cram\'er--von Mises}}
\DeclareAcronym{ML}		    {short={ML},	    	long={maximum likelihood}}
\DeclareAcronym{MLE}		{short={MLE},	    	long={maximum likelihood estimator}}
\DeclareAcronym{MPC}		{short={MPC},	    	long={model predictive control}}
\DeclareAcronym{MPPI}		{short={MPPI},	    	long={model predictive path integral}}
\DeclareAcronym{MSE}		{short={MSE},	    	long={mean squared error}}
\DeclareAcronym{MMSE}		{short={MMSE},	    	long={minimum mean square error}}

\DeclareAcronym{MNR}		{short={MNR},	    	long={matrix norm ratio}}
\DeclareAcronym{MNRE}		{short={MNRE},	    	long={matrix norm relative  error}}
\DeclareAcronym{MSER}		{short={MSER},	    	long={mean squared error ratio}}

\DeclareAcronym{MV}		{short={MV},	    	long={mean variance}}

\DeclareAcronym{NCI}		{short={NCI},	    	long={noncredibility index}}
\DeclareAcronym{NEES}      {short={NEES},	    	long={normalized estimation error squared}}
\DeclareAcronym{NLL}       {short={NLL},	    	long={negative log-likelihood}}
\DeclareAcronym{NGUCE}      {short={NGUCE},	    	long={normalized \ac{GUCE}}}
\DeclareAcronym{NUTS}      {short={NUTS},	    	long={No-U-Turn Sampler}}

\DeclareAcronym{ODE}	    {short={ODE},	    	long={ordinary differential equation}}
\DeclareAcronym{OCP}	    {short={OCP},	    	long={optimal control problem}}

\DeclareAcronym{PBP}		{short={PBP},	    	long={probabilistic backpropagation}}

\DeclareAcronym{PCD}		{short={PCD},	    	long={projected cumulative distribution}}
\DeclareAcronym{PDF}		{short={PDF},	    	long={probability density function}}

\DeclareAcronym{QCE}		{short={QCE},	    	long={quantile calibration error}}

\DeclareAcronym{RL}         {short={RL},            long={reinforcement learning}}
\DeclareAcronym{RMSE}		{short={RMSE},	    	long={root \ac{MSE}}}
\DeclareAcronym{RMV}		{short={RMV},	    	long={root \ac{MV}}}
\DeclareAcronym{RNN}		{short={RNN},	    	long={recurrent neural network}}
\DeclareAcronym{S2KF}	    {short={S$^2$KF},	    long={Smart Sampling Kalman Filter}}
\DeclareAcronym{SVI}		{short={SVI},	    	long={Stochastic Variational Inference}}
\DeclareAcronym{TAGI}		{short={TAGI},	    	long={Tractable Approximate Gaussian Inference}}

\DeclareAcronym{UCE}		{short={UCE},	    	long={uncertainty calibration error}}
\DeclareAcronym{UKF}		{short={UKF},	    	long={Unscented Kalman Filter}}

\DeclareAcronym{VI}		    {short={VI},	    	long={variational inference}}

\title{%
    \LARGE\bf
    Sampling-based Model Predictive Control Using Trust Regions
\author{Markus Walker, Marcel Reith-Braun, Daniel Frisch, and Uwe D. Hanebeck}
\thanks{%
    This work is part of the German Research Foundation (DFG) AI Research Unit 5339 regarding the combination of physics-based simulation with AI-based methodologies for the fast maturation of manufacturing processes.}
\thanks{%
    Markus~Walker, Marcel~Reith-Braun, Daniel~Frisch, and Uwe~D.~Hanebeck are with the Intelligent Sensor-Actuator-Systems Laboratory (ISAS), Institute for Anthropomatics and Robotics, Karlsruhe Institute of Technology, Germany (e-mail:
    \{%
        {\tt\footnotesize firstname.lastname}%
    \}%
    {\tt\footnotesize @kit.edu}%
    ).%
    }
}

\begin{document}

\maketitle
\thispagestyle{empty}
\pagestyle{empty}


\begin{abstract}
Sampling-based \ac{MPC} algorithms, such as \ac{MPPI}, enable approximate, gradient-free solutions to optimal control problems by drawing samples from a proposal distribution, evaluating their trajectory costs, and updating the proposal parameters accordingly.
However, these approaches typically rely on heuristics for adjusting hyperparameters, such as temperature or momentum, or manual tuning.
We propose a trust region formulation for sampling-based \ac{MPC} that constrains updates of the proposal distribution via a principled \ac{KL} divergence bound and, optionally, an entropy lower bound.
This replaces heuristic hyperparameter adaptation with values that are optimal w.r.t. the underlying Lagrangian. 
We further improve sample efficiency and convergence by combining the trust region update with deterministic \ac{LCD}-based sampling.
Experiments on two benchmark environments demonstrate that the proposed trust region update achieves faster convergence and better sample efficiency in low-sample and low-iteration regimes, especially when paired with deterministic \ac{LCD}-based sampling.
\acresetall
\end{abstract}





\section{Introduction}
\label{sec:intro}

Sampling-based \ac{MPC} methods, such as \ac{CEM}--\ac{MPC}~\cite{pinneriSampleefficientCrossentropyMethod2021} and \ac{MPPI}~\cite{williamsInformationtheoretic2018,kazimRecentAdvancesPath2024}, have gained popularity for solving complex \acp{OCP} due to their ability to efficiently parallelize computations and handle arbitrary cost functions and system dynamics.

These methods iteratively refine a parameterized proposal \ac{PDF} over control sequences by drawing samples, evaluating their trajectory costs, and updating the proposal parameters accordingly.
Importance-weighting-based variants such as standard \ac{MPPI}~\cite{williamsInformationtheoretic2018}, and iterative extensions thereof~\cite{bhardwajSTORMIntegratedFramework2022} assign weights to the sampled control sequences based on their associated trajectory costs, thereby using a temperature parameter to control the sharpness of the weighting. 
This parameter is typically held constant or adjusted heuristically during optimization~\cite{pezzatoSamplingbasedModelPredictive2025}.
Similarly, \ac{CEM}-based methods rely on momentum-style smoothing~\cite{rubinsteinCrossEntropy2004} to prevent premature collapse of the proposal distribution.
These heuristics are often difficult to tune and may compromise convergence or performance across different problem settings.

Motivated by the success of trust region methods in \ac{RL}~\cite{schulmanTrustRegionPolicy2015,petersRelativeEntropyPolicy2010}, where principled \ac{KL}-constrained updates have largely replaced heuristic parameter tuning, we propose to transfer these ideas to sampling-based \ac{MPC}.
Specifically, for each proposal update, we constrain the new proposal to remain within a bounded divergence from the previous one. 
The resulting constrained optimization problem is solved via Lagrange multipliers, with optimal hyperparameters obtained from the dual formulation. 
The updated proposal is then projected back onto the original distribution family (\eg, Gaussian), yielding a feasible distribution from which control sequences are sampled in the subsequent iteration.

Orthogonal to the choice of update rule, the sampling method directly affects the quality and smoothness of the control.
In our previous work~\cite{walkerSampleefficientSmoothCrossentropy2026,arXiv26_Walker_IFAC}, we demonstrated that exchanging random samples with precomputed, optimally-placed deterministic samples based on \acp{LCD}~\cite{CDC09_HanebeckHuber} yield improved sample efficiency and smoother control sequences.
\ac{LCD}-based sampling shows superior coverage and inherently positions the samples homogeneously, \ie, it induces a smooth distribution of the sizes of the sample-free gaps between them.
It is therefore well-suited for integration and optimization tasks.
In particular, it integrates well with trust region updates, which restrict the proposal to remain close to the original and thus to regions with high sample coverage.


%
%
\paragraph*{Contribution}

Based on the success of trust region methods in \ac{RL}, we propose a trust region formulation for sampling-based \ac{MPC} that replaces heuristic proposal adaptations with principled \ac{KL}-constrained updates derived from dual optimization.
Building on our previous work~\cite{walkerSampleefficientSmoothCrossentropy2026,arXiv26_Walker_IFAC}, we further combine the trust region update with deterministic \ac{LCD}-based sampling to improve sample efficiency.
Both contributions are evaluated on two benchmark control tasks and compared against heuristic baselines across multiple sampling strategies, including random and low-discrepancy sampling (Sobol~\cite{joeConstructingSobolSequences2008} and Halton~\cite{owen2017randomizedhaltonalgorithmr}).

\paragraph*{Notation}
In this paper, underlined letters, \eg, $\vx$, denote vectors, boldface letters, such as $\rvx$, represent random variables, while boldface capital letters, such as $\mA$, indicate matrices.
The expectation of a random variable is denoted by $\hat{\cdot}$, \eg, $\evx$, and covariance matrices are denoted by $\mC$.
\section{Background}
\label{sec:ocp}

We consider a finite-horizon \ac{OCP} with cumulative cost
\begin{align}
    \label{eq:ocp_cost}
      J_k & = g_{\horizon}(\vx_{k + \horizon}) + \sum_{n=0}^{\horizon-1} g_n(\vx_{k + n}, \vu_{k + n})  \enspace,
\end{align}
where $k$ denotes the current time step, $\horizon$ the prediction horizon, and $n$ the time step within the prediction horizon.
Furthermore, $\vx_k \in \IR^{d_x}$ represents the system state, and $\vu_k \in \IR^{d_u}$ the control input.
The stage cost is given by $g_n(\cdot, \cdot)\colon \IR^{d_x} \times \IR^{d_u} \to \IR$, and the terminal cost by $g_{\horizon}(\cdot)\colon \IR^{d_x} \to \IR$.
The discrete-time system dynamics are described by
\begin{align}
    \label{eq:system_dynamics}
    \vx_{n+1} & = \va_n(\vx_n, \vu_n) \enspace, \quad n=k,\ldots,k+\horizon-1 \enspace,
\end{align}
with transition function $\va_k(\cdot,\cdot)\colon \IR^{d_x} \times \IR^{d_u} \to \IR^{d_x}$.

The objective is to determine the optimal control sequence $\vu_{k:k+\horizon-1}^* = \left(\vu_k^*, \vu_{k+1}^*, \ldots, \vu_{k+\horizon-1}^*\right)$ that minimizes $J_k$.
From this optimal sequence, the first control input $\vu_k^*$ is applied to the system, and the optimization is repeated at the next time step $k+1$ in a receding horizon fashion.
For compact notation, we represent the stacked control sequence $\vu_{k:k+\horizon-1}$ by $\vVarCem \in \IR^{\dimVarCem}$, \ie, $\vVarCem = [\vu_k\T, \vu_{k+1}\T, \ldots, \vu_{k+\horizon-1}\T]\T \in \IR^{\horizon \cdot d_u}$ (omitting time index $k$ from now on for brevity) and write the associated cumulative cost as $J(\vVarCem)$.

In sampling-based \ac{MPC}, this \ac{OCP} is solved iteratively by drawing $\numCEMSamples$ control sequences $\{\vVarCem^{(i)}\}_{i=1}^{\numCEMSamples}$ from a proposal \ac{PDF} $\pdf(\vVarCem; \vtheta_j)$, parameterized by $\vtheta_j$ at iteration $j=0,\dots, \numCEMIter-1$. 
For each sampled control sequence $\vVarCem^{(i)}$, the corresponding state trajectory is predicted using the system dynamics, its cost is evaluated, and the proposal is subsequently updated based on all trajectories and their costs.
A common choice for the proposal in the considered Euclidean action space is a Gaussian distribution $\Gaussian(\vVarCem; \evVarCem_j, \mC_j)$, where the mean $\evVarCem_j$ and covariance $\mC_j$ constitute the parameters collected in $\vtheta_j$.

\section{Related Work}
\label{sec:related_work}

\subsection{Sampling-based \acs*{MPC}}

A prominent approach to sampling-based \ac{MPC} is the \ac{CEM}, which iteratively updates a proposal \ac{PDF} by selecting low-cost elite samples and maximizing their likelihood~\cite{rubinsteinCrossEntropy2004}.
For Gaussian proposals, the updated moments for the next iteration are computed as the empirical mean and covariance of the selected control sequences~\cite{rubinsteinCrossEntropy2004}.
To improve numerical robustness and reduce premature concentration of the proposal, many implementations introduce a momentum (or smoothing) term across iterations~\cite{rubinsteinCrossEntropy2004,pinneriSampleefficientCrossentropyMethod2021}.
For the mean update, this is typically written as
\begin{align}
    \label{eq:cem_momentum_mean}
    \evVarCem_{j+1} = \alpha \, \evVarCem_{j} + (1-\alpha) \, \evVarCem_{j}^{\prime} \enspace,
\end{align}
where $\evVarCem_{j}^{\prime}$ denotes the (unsmoothed) mean of the elite set.

Related \ac{MPPI}-like methods also perform (iterative) proposal refinement, but they rely on importance weighting of sampled control sequences instead of hard elite selection~\cite{arXiv26_Walker_IFAC}.
The weights are given by
\begin{align}
    \label{eq:imp_weights}
    w^{(i)} \propto \exp\left(-\frac{1}{\lambda} J\left(\vVarCem^{(i)}\right) \right) \enspace,
\end{align}
and are used to compute the updated Gaussian moments as the control sequences' weighted sample mean and covariance.
The inverse temperature $\lambda$ controls the exploration-exploitation trade-off and remains constant~\cite{williamsInformationtheoretic2018} or is adapted heuristically during optimization~\cite{pezzatoSamplingbasedModelPredictive2025}.
Furthermore, iterative \ac{MPPI} variants additionally employ momentum-style smoothing~\cref{eq:cem_momentum_mean} of the proposal update, analogous to \ac{CEM}, to mitigate premature convergence~\cite{bhardwajSTORMIntegratedFramework2022}.

Orthogonal to the above methods, improvements in sampling efficiency and smoothness of the optimized control inputs have been achieved by using low-discrepancy sequences, such as the Halton sequence~\cite{bhardwajSTORMIntegratedFramework2022}, time-correlated noise~\cite{pinneriSampleefficientCrossentropyMethod2021}, or, as in our previous work, by using optimal deterministic samples based on the \ac{LCD}~\cite{walkerSampleefficientSmoothCrossentropy2026,arXiv26_Walker_IFAC}.

\subsection{Trust Regions in \ac*{RL}}
In \ac{RL}, trust region methods provide principled policy updates that balance exploration and exploitation while ensuring stable learning.
The core idea is to limit the update step by constraining the \ac{KL} divergence between the new policy and the old policy, preventing drastic changes that can lead to performance degradation.
Adding an entropy constraint further encourages exploration and prevents premature convergence to suboptimal policies.
These concepts have been established for policy search~\cite{petersRelativeEntropyPolicy2010,abdolmaleki2015model}, step-based \ac{RL}~\cite{schulmanTrustRegionPolicy2015}, episodic \ac{RL}~\cite{ottoDeepBlackboxReinforcement2023}, and gradient-based stochastic optimal control with quadratic reward functions~\cite{blessingTrustRegionConstrained2025}.
\section{Trust Region \ac*{MPC}}
\label{sec:tr_mpc}

Inspired by the success of trust region methods in \ac{RL}, we propose a trust region formulation for sampling-based \ac{MPC} to improve the optimization process and remove the need for heuristic parameter adaptation.

At iteration $j$, let $\pdf_j$ denote the current proposal \ac{PDF} over control sequences $\vVarCem$, and $\pdf_{j+1}$ the next proposal \ac{PDF} to be optimized.
The trust region optimization problem is then given by
\begin{align}
    \label{eq:tr_objective_function}
    &\min_{\pdf_{j+1}} \Eop_{\pdf_{j+1}}\{J(\vVarCem)\}
    \quad \text{s.t.} \quad D_{\mathrm{KL}}(\pdf_{j+1} \| \pdf_j) \leq \epsilon \enspace, 
    \\
    &
    \quad H(\pdf_{j+1}) \ge H_{\min} \enspace,
    \quad  \Eop_{\pdf_{j+1}} \{ 1\} = 1 \enspace,
\end{align}
where $D_{\mathrm{KL}}(\cdot \| \cdot)$ is the \ac{KL} divergence measuring the distributional change, $H(\cdot)$ is the entropy, $\epsilon>0$ is the maximum allowed \ac{KL} divergence between consecutive proposals,  $H_{\min}$ is the minimum required entropy of the proposal, and the last constraint ensures that $\pdf_{j+1}$ is normalized.
The \ac{KL} constraint prevents overly aggressive updates, while the minimum entropy constraint encourages exploration and mitigates premature collapse of the proposal.
Note that the entropy constraint is optional and can be omitted if task-specific knowledge suggests that proposal collapse is not a concern, \eg, when a small number of iterations is used.

To solve the constrained problem, we form the Lagrangian 
\begin{align}
    & \mathcal{L}(\pdf_{j+1},\eta,\alpha, \nu) = \Eop_{\pdf_{j+1}}\{J(\vVarCem)\}
    + \eta\left(D_{\mathrm{KL}}(\pdf_{j+1} \| \pdf_j) - \epsilon\right) 
    \\ & \quad
    \label{eq:tr:lagrangian}
    + \alpha\left(H_{\min} - H(\pdf_{j+1})\right) 
    + \nu \left( \Eop_{\pdf_{j+1}} \{ 1\} - 1 \right)
    \enspace,
\end{align}
where $\eta$, $\alpha$, and $\nu$ are the Lagrange multipliers associated with the \ac{KL} divergence, entropy, and normalization constraints, respectively.
To find the optimal proposal for the next iteration, we take the functional derivative of $\mathcal{L}$ w.r.t. $\pdf_{j+1}$ and set it to zero (see \cref{sec:appendix:lagrangian} for details).
The optimal proposal is then given by
\begin{align}
    \label{eq:opt_proposal}
    \pdf_{j+1}^{\ast}(\vVarCem) &= \frac{1}{Z_{\eta,\alpha}}\; \pdf_j(\vVarCem)^{\frac{\eta}{\eta+\alpha}}\exp\left(-\frac{J(\vVarCem)}{\eta+\alpha}\right)
    \enspace,
    \\
    \label{eq:opt_proposal:normalization}
    Z_{\eta,\alpha} 
    &= \Eop_{\pdf_j}\left\{ \pdf_j(\vVarCem)^{\tfrac{\eta}{\eta+\alpha}-1}\exp \left( -\frac{J(\vVarCem)}{\eta+\alpha} \right) \right\} \enspace,
\end{align}
where $Z_{\eta,\alpha}$ is the normalization constant, written as expectation over $\pdf_j$.
Since the multiplier $\nu$ is implicitly handled in the normalization, it can be omitted from subsequent derivation steps.
Note that in the case of a Gaussian proposal $\pdf_j(\vVarCem)$, for $\alpha>0$, $\pdf_{j+1}^{\ast}$ is generally non-Gaussian.
However, for $\alpha=0$, and convex quadratic $J(\vVarCem)$, Gaussianity is preserved.

By substituting $\pdf_{j+1}^{\ast}$ back into the simplified Lagrangian $\mathcal{L}(\pdf_{j+1}^{\ast},\eta,\alpha)$ (see \cref{sec:appendix:dual}), we obtain the dual function
\begingroup
\begin{align}
    \label{eq:tr_dual}
    g(\eta, \alpha) 
    = -\eta \epsilon + \alpha H_{\min} - (\eta + \alpha) \log Z_{\eta,\alpha} \enspace,
\end{align}
\endgroup
which we use to find the optimal Lagrange multipliers $\eta^{\ast}$ and $\alpha^{\ast}$ by numerically solving the dual problem
\begin{align}
    \label{eq:tr_dual:opt_problem}
    \max_{\eta \geq 0, ~\alpha \geq 0} g(\eta, \alpha) \enspace.
\end{align}
Note that, in our case, strong duality holds, so that solving the dual problem is equivalent to solving the primal problem~\cref{eq:tr_objective_function} (see \cref{sec:appendix:strong_duality}) and that $\pdf_{j+1}$ never degenerates, \ie, $\eta^{\ast} + \alpha^{\ast} > 0$ (see \cref{sec:appendix:pos_multiplier_sum}).

Since the optimization over $\eta$ and $\alpha$ is always two-dimensional (in particular, it is independent of the control input dimension and horizon length), standard gradient-based solvers are efficient in practice.
The gradient of the dual function $\nabla g(\eta,\alpha)$ is given by
\begin{align}
    \label{eq:tr_dual:gradient}
    \begin{bmatrix}
        \frac{\partial g}{\partial \eta}
        \\
        \frac{\partial g}{\partial \alpha}
    \end{bmatrix}
    =
    \begin{bmatrix}
        -\log Z_{\eta,\alpha}
        -(\eta+\alpha)\frac{\partial \log Z_{\eta,\alpha}}{\partial \eta}
        -\epsilon
        \\
        -\log Z_{\eta,\alpha}
        -(\eta+\alpha)\frac{\partial \log Z_{\eta,\alpha}}{\partial \alpha}
        +H_{\min}
    \end{bmatrix}
\end{align}
where the partial derivatives of $\log Z_{\eta,\alpha}$ are
\begin{align}
    &
    \frac{\partial \log Z_{\eta,\alpha}}{\partial \eta}
    =
    \frac{1}{(\eta+\alpha)^2}
    \\
    &
    \enspace\cdot
    \frac{\Eop_{\pdf_{j}} \left\{
    \pdf_{j}(\vVarCem)^{\frac{\eta}{\eta+\alpha}-1}
    \exp \left(-\frac{J(\vVarCem)}{\eta+\alpha}\right)
    \left(\alpha\log \pdf_{j}(\vVarCem)+J(\vVarCem)\right)
    \right\} }
    {
        Z_{\eta,\alpha}
    } \, ,
    \\
    &
    \frac{\partial \log Z_{\eta,\alpha}}{\partial \alpha}
    =
    \frac{1}{(\eta+\alpha)^2}
    \\
    &
    \enspace\cdot
    \frac{\Eop_{\pdf_{j}} \left\{
    \pdf_{j}(\vVarCem)^{\frac{\eta}{\eta+\alpha}-1}
    \exp \left(-\frac{J(\vVarCem)}{\eta+\alpha}\right)
    \left(-\eta\log \pdf_{j}(\vVarCem)+J(\vVarCem)\right)
    \right\} }
    {
        Z_{\eta,\alpha}
    } \, .
\end{align}
Note that the gradient is entirely written in terms of expectations of $\pdf_j$.
Consequently, the gradient can be estimated from samples drawn from $\pdf_j$.



Once the optimal Lagrange multipliers $\eta^{\ast}$ and $\alpha^{\ast}$ have been obtained, the optimal proposal \ac{PDF} $\pdf_{j+1}^{\ast}$ follows directly from~\cref{eq:opt_proposal}.
We can derive a Dirac-mixture approximation of $\pdf_{j+1}^{\ast}$ based on samples drawn from $\pdf_j$ and weights
\begin{align}
    \label{eq:tr_dual:weights}
    w_i \propto \pdf_j(\vVarCem^{(i)})^{\frac{\eta^{\ast}}{\eta^{\ast}+\alpha^{\ast}}-1}\exp\left(-\frac{J(\vVarCem^{(i)})}{\eta^{\ast}+\alpha^{\ast}}\right) \enspace.
\end{align}
Finally, we project our Dirac-mixture approximation of $\pdf_{j+1}^{\ast}$ onto the same parametric form as the original proposal \ac{PDF}, \eg, via moment matching for Gaussians, or expectation--maximization for Gaussian mixtures.
In this work, we use Gaussian proposals, which yield a closed-form update for the new mean $\evVarCem_{j+1}$ and covariance $\mC_{j+1}$, expressed in terms of the weighted empirical sample mean and covariance.
Notably, when $\alpha^{\ast} = 0$, this projection does not introduce additional errors, as the true $\pdf_{j+1}^{\ast}$ is Gaussian. 
In the general case, however, the projection incurs an additional approximation error.

Although different in derivation, the resulting weights \cref{eq:tr_dual:weights} exhibit a form similar to those in \cref{eq:imp_weights} used by \ac{MPPI}-like methods.
In particular, when the entropy constraint is inactive, \ie, $\alpha^{\ast}=0$, the two formulations coincide exactly.
Importantly, however, our approach yields optimal weights in a principled manner from the trust region formulation, thereby eliminating the need for heuristic hyperparameter adaptation.
\section{Algorithmic Improvements}
\label{sec:alg_improvements}

Beyond the trust region formulation, we introduce several algorithmic improvements.
An overview is provided in \cref{alg:tr_mpc}, where the additional steps are highlighted in {\color{myblue}blue}.
Among these, the sampling method is a central design choice: we use random sampling as a baseline and introduce low-discrepancy and deterministic \ac{LCD}-based sampling as an extension.

\begin{algorithm}[t]
    \small
    \caption{Trust Region MPC Step}
    \label{alg:tr_mpc} 
    \SetKwFunction{FMain}{TR-MPC-Step}
    \SetKwInOut{Input}{Input}
    \Input{%
        State $\vx_k$, initial parameters $\paramCEM_0 = (\evVarCem_0, \mC_0)$, {\color{myblue} buffered samples from previous step}
    }%
    \SetKwInOut{Output}{Optimal control input $\vu_k^*$}
    {\color{myblue} Warm start last proposal mean and buffered samples}\;
    \For{$j \gets 0$ \KwTo $\numCEMIter-1$}{
        Sample $\vu^{(1)}_{k:k+\horizon-1}, \ldots, \vu^{(\numCEMSamples)}_{k:k+\horizon-1} \sim \pdf(\cdot; \paramCEM_j)$\; \label{alg:tr_mpc:sampling_step}
        {\color{myblue} Add saved samples from buffer}\;
        {\color{myblue} Clip samples to input bounds}\;
        Trajectory shooting using \cref{eq:system_dynamics}\tcp*{\small \tt parallel}
        Evaluate costs $\{J_k(\vu^{(i)}_{k:k+\horizon-1})\}_{i=1}^{\numCEMSamples}$ using~\cref{eq:ocp_cost}\tcp*{\tt\small parallel}
        Solve dual problem~\cref{eq:tr_dual:opt_problem}\;
        Update proposal parameters $\paramCEM_{j+1}$ (weighted sample mean and covariance) using \cref{eq:tr_dual:weights} \;
        {\color{myblue} Add $\bufferSize$ best samples to buffer}\;
    }
    \Return first control $\vu_k^*$ from best sampled sequence
\end{algorithm}

\subsection{Sampling Methods}
\label{sec:alg_improvements:lcd_samples}
We propose to use deterministic samples based on \acp{LCD}~\cite{CDC09_HanebeckHuber} to represent the isotropic standard normal distribution $\mathcal{N}(\vVarCem;\vzero,\mI)$ optimally with a fixed number of samples. 
The optimization is done \emph{offline}, and these samples are stored for use in optimization iterations. 
For this, we use our Python package\footnote{\url{https://github.com/KIT-ISAS/deterministic_gaussian_sampling_py}}.

\begin{figure}[t]
    \centering
    \begin{subfigure}{0.49\columnwidth}
        \includegraphics{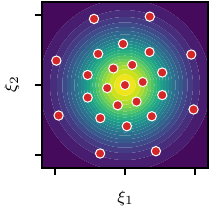}%
        \caption{Standard normal samples}%
        \label{fig:lcd_samples:standard}%
    \end{subfigure}
    \begin{subfigure}{0.49\columnwidth}
        \includegraphics{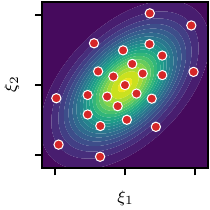}%
        \caption{Transformed samples}%
        \label{fig:lcd_samples:transformed}%
    \end{subfigure}%
    \caption{Example showing \num{25} two-dimensional deterministic samples, where the \ac{PDF} is indicated by the background color (adapted  from~\cite{walkerSampleefficientSmoothCrossentropy2026}.)}%
    \label{fig:lcd_samples}%
\end{figure}

At each iteration $j$ the stored samples are mapped to match the proposal parameters $\paramCEM_j=(\evVarCem_j,\mC_j)$ via
\begin{align}
    \label{eq:lcd_sample_transform}
    \vVarCem_j^{(i)} = \evVarCem_j + \mL_j \, \mR_j \, \vVarLcd_{\mathrm{SN}}^{(i)} \enspace,
\end{align}
where $\mL_j$ satisfies $\mC_j=\mL_j\mL_j\T$. 
A random rotation matrix $\mR_j$ is drawn at each iteration from the special orthogonal group $SO(\dimVarCem)$ (orthogonal matrices with determinant $+1$)~\cite{leonStatisticalModelRandom2006}. 
Applying the rotation before scaling preserves the optimality because the standard normal distribution is rotationally invariant, while the random rotation introduces stochasticity that improves exploration~\cite{walkerSampleefficientSmoothCrossentropy2026}.
\Cref{fig:lcd_samples} illustrates the precomputed sample set and the transformation.

As alternatives, we also consider scrambled (randomized) low-discrepancy sampling using Sobol~\cite{joeConstructingSobolSequences2008} and Halton~\cite{owen2017randomizedhaltonalgorithmr} sequences, as well as standard random sampling.
Note that the root mean square error of random sampling for approximating expectations (integration) decreases proportionally to $\numCEMSamples^{-\frac{1}{2}}$, while low-discrepancy sampling scales proportionally to $(\log \numCEMSamples)^{\dimVarCem}\numCEMSamples^{-1}$~\cite{Dick2013_HighDimensionalIntegrationTheQuasiMonteCarloWay}.

In comparison to low-discrepancy sampling, \ac{LCD}-based sampling has the advantage of allowing deterministic samples to be optimized directly for Gaussians. 
In contrast, low-discrepancy sampling, such as Halton and Sobol, requires a transformation from a uniform distribution to a Gaussian distribution, which can introduce additional error and complexity.

\subsection{Standard Improvements}
\label{sec:standard_improvements}

Besides sampling methods, we apply several standard enhancements commonly used in sampling-based \ac{MPC}.
We use warm starting by initializing the proposal mean at iteration $j = 0$ with the shifted mean from the previous time step, in which the last control input is duplicated to fill the horizon.
To enforce control input bounds, sampled controls are clipped to the admissible range. This can be viewed as modified system dynamics incorporating an additional saturation nonlinearity~\cite{williamsInformationtheoretic2018}.

We further use a buffer of size $\bufferSize$ that retains the $\bufferSize$ best trajectories (cost-based ranking) from the previous iteration, as in~\cite{walkerSampleefficientSmoothCrossentropy2026,arXiv26_Walker_IFAC}, to improve sample efficiency. 
As with the mean of the proposal, the buffered samples are also warm-started in iteration $j = 0$ in the same way as described above.

Finally, to encourage smooth control trajectories, we use a time-correlated proposal covariance.
The time-correlation structure is derived from colored noise with power spectral density $\mathrm{PSD}(\omega) \propto \nicefrac{1}{\omega^{\beta}}$, where $\beta$ controls the noise color and $\omega$ denotes frequency. Within a time step, the controls are assumed to be uncorrelated.
During optimization, only marginal variances are updated, while the correlation structure is kept fixed, since estimating full covariance matrices from a few samples is error-prone.
Further details are provided in~\cite{arXiv26_Walker_IFAC}.
\section{Experiments}
\label{sec:eval}

We evaluate the proposed method on two benchmark tasks: (i) cart-pole swing-up and (ii) truck backer-upper.
For the cart-pole swing-up task, the initial pole angle $\phi$ is sampled uniformly from $[145^\circ, 215^\circ]$, where $\phi = 0^\circ$ corresponds to the upright position.
For the truck backer-upper task, the initial $x$ and $y$ positions are sampled uniformly from $[15, 20]$ and $[-10, 10]$, respectively. The trailer and truck orientations are drawn uniformly from $[-90^\circ, 90^\circ]$, with the goal of backing the trailer to the position $(x,y)=(0,0)$.
Detailed simulation and controller settings for these tasks can be found in~\cite{arXiv26_Walker_IFAC,walkerSampleefficientSmoothCrossentropy2026}.
To assess the method beyond these classical tasks, we additionally consider (iii) CheetahRun and (iv) WalkerRun from MuJoCo Playground~\cite{zakkaDemonstratingMuJoCoPlayground2025} as supplementary validation.

We first analyze the effect of the trust region parameters and the proposal-update behavior, including the entropy lower bound, before comparing the resulting method against heuristic \ac{MPPI}-like baselines.
As baseline, we use an \ac{MPPI}-like method from~\cite{arXiv26_Walker_IFAC}, which also updates proposal mean and covariance from weighted samples but does not enforce a trust region. Instead, it relies on a temperature heuristic~\cite{pezzatoSamplingbasedModelPredictive2025} and momentum-style smoothing~\cref{eq:cem_momentum_mean} to stabilize optimization.

To isolate the effect of the trust region formulation, both methods use the same algorithmic enhancements from \cref{sec:alg_improvements} and the same hyperparameters as in~\cite{arXiv26_Walker_IFAC}.
We compare performance in terms of sample efficiency, smoothness of the control inputs, and convergence over iterations.
Smoothness is measured using $\sum_{k=1}^{T-1} \| \vu_{k} - \vu_{k-1} \|^2 $, where $T$ is the number of time steps in a simulation (see~\cite{arXiv26_Walker_IFAC}), and a lower value indicates a smoother control trajectory.
Both sample efficiency and convergence are evaluated in terms of cumulative cost, which is the sum of stage costs over $T$ steps.

We consider four sampling methods: random sampling, low-discrepancy sampling (Sobol~\cite{joeConstructingSobolSequences2008} and Halton~\cite{owen2017randomizedhaltonalgorithmr}), and the deterministic \ac{LCD}-based sampling. 
To transform the uniform Sobol and Halton samples to an arbitrary Gaussian, we use the inverse transform and eigenvalue decomposition, as described in~\cite{frischGeneralizedFibonacciGrid2023}, with no further moment correction.
In the plots, heuristic baseline variants are shown with solid lines, and trust region variants are shown with dashed lines labeled with suffix \emph{TR}. 
Labels indicate the sampling method, where the prefix \emph{ds} denotes deterministic \ac{LCD} sampling. 
Furthermore, we compare to the \ac{CEM} using \ac{LCD}-based samples~\cite{walkerSampleefficientSmoothCrossentropy2026} denoted by \emph{dsCEM}.

\begin{figure}%
    \centering%
    \begin{subfigure}{\columnwidth}
        \includegraphics[width=\linewidth,trim=0 0 0.05cm 0.1cm, clip]{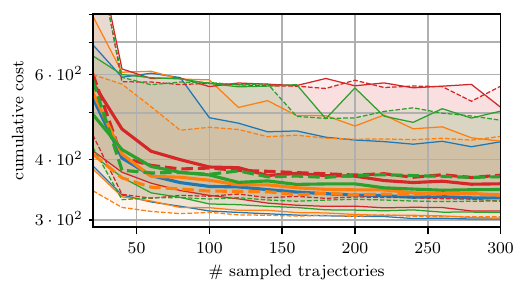}%
        \caption{Cumulative costs over sample size}%
        \label{fig:cart_pole:cum_cost}%
    \end{subfigure}
    \begin{subfigure}{\columnwidth}
        \includegraphics[width=\linewidth,trim=0 0 0.05cm 0.1cm, clip]{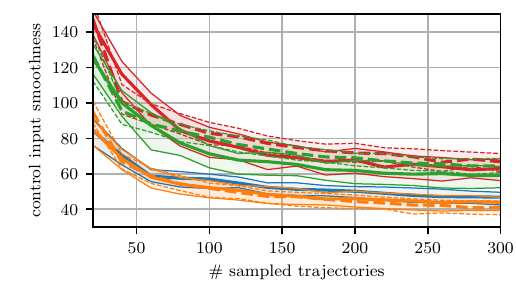}%
        \caption{Control input smoothness}%
        \label{fig:cart_pole:action_smoothness}%
    \end{subfigure}
    \begin{subfigure}{\columnwidth}
        \centering%
        \includegraphics[width=\linewidth, ,trim=0 0 0.05cm 0.1cm, clip]{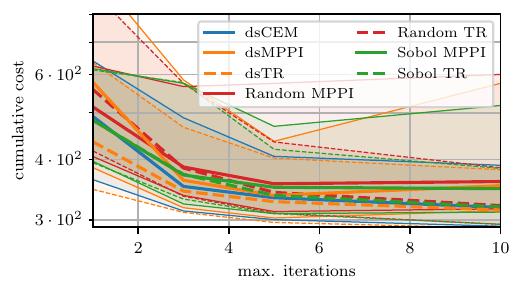}%
        \caption{Cumulative costs over optimizer iterations}%
        \label{fig:cart_pole:convergence}%
    \end{subfigure}    
    \caption{Results for the cart-pole swing-up environment. All plots share the same legend.
    }
    \label{fig:cart_pole:results}%
\end{figure}

\begin{figure}%
    \centering%
    \begin{subfigure}{\columnwidth}
        \includegraphics[width=\linewidth]{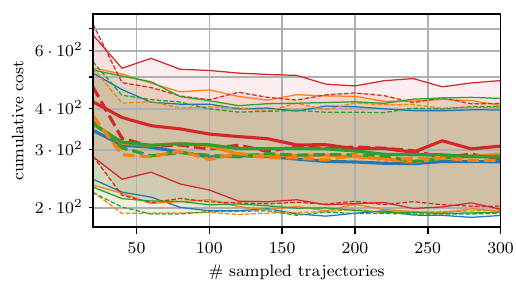}%
        \caption{Cumulative costs over sample size}%
        \label{fig:truck:cum_cost}%
    \end{subfigure}
    \begin{subfigure}{\columnwidth}
        \includegraphics[width=\linewidth]{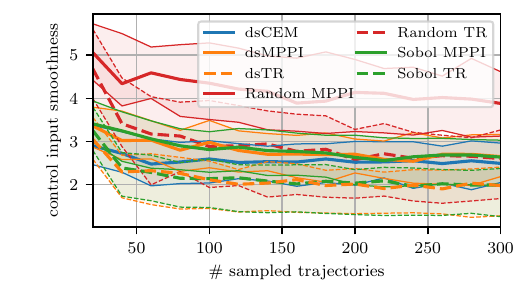}%
        \caption{Control input smoothness}%
        \label{fig:truck:action_smoothness}%
    \end{subfigure}
    \begin{subfigure}{\columnwidth}
        \centering%
        \includegraphics[width=\linewidth]{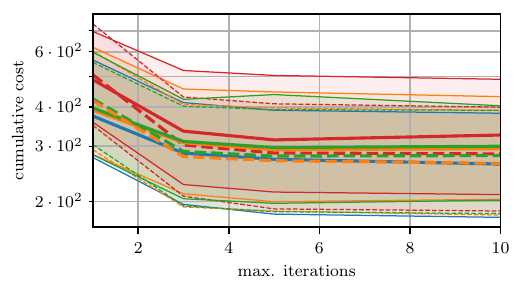}%
        \caption{Cumulative costs over optimizer iterations}%
        \label{fig:truck:convergence}%
    \end{subfigure}    
    \caption{Results for the truck backer-upper environment.
    All plots share the same legend as in \cref{fig:cart_pole:results}.
    }
    \label{fig:truck:results}%
\end{figure}

Each method and setting is evaluated over \num{100} runs, and statistics are reported as median and interquartile range. 
Results are shown in \cref{fig:cart_pole:results,fig:truck:results,fig:eps_sweep}.

\subsection{Trust Region Parameter Sensitivity and Proposal Behavior}

\begin{figure}
    \centering
        \includegraphics[width=\linewidth]{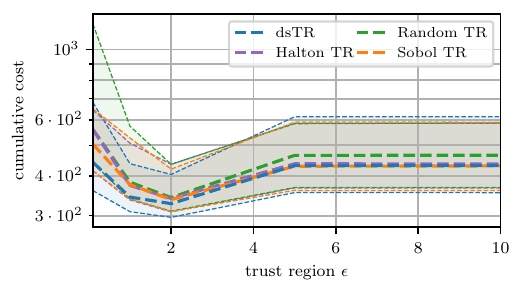}%
    \caption{Cumulative costs over trust region parameter $\epsilon$ for the cart-pole swing-up environment.
    }
    \label{fig:eps_sweep}%
\end{figure}

The $\epsilon$-sweep result for the proposed trust region method with random, low-discrepancy, and deterministic \ac{LCD} sampling is shown in \cref{fig:eps_sweep} for \num{100} samples and \num{5} optimizer iterations, exemplified for cart-pole swing-up.
The lowest cumulative cost is achieved at $\epsilon = 2$ for all sampling methods.
If $\epsilon$ is larger, the trust region constraint is looser and allows for more aggressive updates that can lead to worse performance. 
When $\epsilon$ is smaller, the trust region constraint is tighter, which can lead to more conservative updates requiring more iterations to achieve good performance.
However, if large iteration counts are feasible, a smaller $\epsilon$ can achieve better performance.

\begin{figure}%
    \centering%
    \begin{subfigure}{0.48\columnwidth}
        \includegraphics[, trim=0 0 1.3cm 0.0cm, clip]{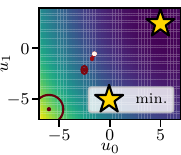}%
        \caption{dsCEM}%
        \label{fig:cart_pole:100step:dscem}%
    \end{subfigure}
    \begin{subfigure}{0.48\columnwidth}
        \includegraphics[, trim=0 0 0 0.0cm, clip]{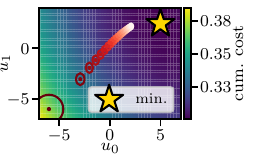}%
        \caption{dsMPPI}%
        \label{fig:cart_pole:100step:dsmppi}%
    \end{subfigure}
    \\
    \begin{subfigure}{0.48\columnwidth}
        \includegraphics[, trim=0 0 1.3cm 0.0cm, clip]{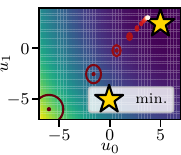}%
        \caption{dsTR (no entropy constraint)}%
        \label{fig:cart_pole:100step:dstr_no_entropy}%
    \end{subfigure}
    \begin{subfigure}{0.48\columnwidth}
        \centering%
        \includegraphics[, trim=0 0 0 0.0cm, clip]{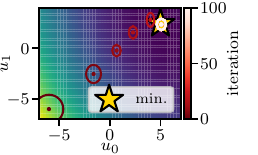}%
        \caption{dsTR (entropy constrained)}%
        \label{fig:cart_pole:100step:dstr_entropy_constrained}%
    \end{subfigure}    
    \caption{Convergence of a single cart-pole control step with $\horizon=\num{2}$ ($\vxi=[u_1,u_2]\T$), $\epsilon=2$, and $H_{\min}=0$ for the pathological case where the initial proposal is far from $\min J(\vxi)$. Gaussian proposals with diagonal covariance are shown as $2\sigma$ ellipses over \num{100} iterations; every 10th iterate is plotted and color indicates the iteration number. Without the entropy constraint (a--c), the proposal collapses prematurely; with it, it retains exploration and converges to the optimum.
    }
    \label{fig:cart_pole:100step}%
\end{figure}

The entropy lower bound $H_{\min}$ has little effect in the settings considered in the main benchmark comparisons, where the proposal does not collapse within the feasible iteration counts. 
However, for longer optimization runs, $H_{\min}$ prevents premature collapse of the proposal and preserves exploration, as shown in \cref{fig:cart_pole:100step}. Without the entropy constraint, the proposal collapses and fails to reach the global minimum; with the entropy constraint, it retains sufficient exploration to converge toward the optimum.

For the main comparisons, the trust region parameters are fixed to $\epsilon = 2$ and $H_{\min} = -50$.
Since Sobol and Halton samples show very similar trends, we present Sobol only in the following plots for readability.

\subsection{Sample Efficiency}
\label{sec:eval:sample_efficiency}

In the sample-efficiency analysis, the number of optimizer iterations is fixed to three, and sample sizes from \num{20} to \num{300} are evaluated (see \cref{fig:cart_pole:cum_cost,fig:truck:cum_cost}).
In both environments, trust region variants mostly achieve lower cumulative costs than heuristic variants when the sample budget is small (e.g., $N \le 100$),  which indicates improved sample efficiency.
%
Across settings, the trust region method with deterministic \ac{LCD} sampling consistently outperforms its heuristic counterpart and dsCEM.
This advantage is particularly pronounced for the cart-pole swing-up, where the proposed dsTR clearly achieves the best performance in the low-sample regime.

\subsection{Control Input Smoothness}
\label{sec:eval:smoothness}
The smoothness comparison, shown in \cref{fig:cart_pole:action_smoothness,fig:truck:action_smoothness}, uses the same sample and iteration settings as in \cref{sec:eval:sample_efficiency}.
In the truck backer-upper environment, trust region variants improve control-input smoothness over their heuristic counterparts at low sample sizes across all sampling strategies.
In cart-pole swing-up, heuristic variants with random and Sobol sampling are smoother than their corresponding trust region variants for most sample sizes.
For deterministic \ac{LCD} sampling, however, dsTR achieves slightly lower or equal smoothness values than dsMPPI and dsCEM in most settings.
Therefore, trust region updates do not necessarily improve smoothness, and their smoothness mainly depends on the sampling strategy.

\subsection{Convergence}

In the convergence analysis, displayed in \cref{fig:cart_pole:convergence,fig:truck:convergence}, the sample size is fixed to $N=100$, and the maximum number of iterations ($\numCEMIter$) is varied over $\{1,3,5,10\}$.
Trust region variants show faster convergence than heuristic variants and dsCEM, with lower median cumulative costs and tighter interquartile ranges across iterations.
This effect is particularly pronounced for the truck backer-upper: even after ten iterations, heuristic variants do not reach the cost achieved by trust region variants using only three iterations for any sampling strategy.
Across both environments, dsTR provides the best convergence behavior, achieving lower costs with fewer iterations than all alternatives.

\subsection{Runtime}

\begin{table}[t]
    \centering    
    \caption{Computation times per control step ($\numCEMIter=3$).}
    \label{tab:computation_times}
    \setlength{\tabcolsep}{2pt}
    {
    \small
    \sisetup{round-mode=places,round-precision=1, scientific-notation=fixed, fixed-exponent=-3, drop-exponent=true}
    \begin{tabular}{lcc}
        \toprule
        Method      & Cart-Pole Swing-Up$^{\text{a}}$             & Truck Backer-Upper$^{\text{a}}$  
        \\
        \midrule
        dsMPPI      & $\num{0.1710} \pm \num{0.0469}$ & $\num{0.0691} \pm \num{0.0422}$
        \\
        dsTR        & $\num{0.2018} \pm \num{0.0557}$ & $\num{0.0836} \pm \num{0.0472}$                  
        \\
        Halton MPPI & $\num{0.1713} \pm \num{0.0414}$ & $\num{0.0806} \pm \num{0.0467}$
        \\
        Halton TR   & $\num{0.2086} \pm \num{0.0524}$ & $\num{0.0980} \pm \num{0.0530}$
        \\
        Sobol MPPI  & $\num{0.1840} \pm \num{0.0519}$ & $\num{0.0832} \pm \num{0.0473}$
        \\
        Sobol TR    & $\num{0.2118} \pm \num{0.0558}$ & $\num{0.0955} \pm \num{0.0517}$
        \\
        Random MPPI & $\num{0.1661} \pm \num{0.0488}$ & $\num{0.0645} \pm \num{0.0364}$
        \\
        Random TR   & $\num{0.1995} \pm \num{0.0504}$ & $\num{0.0789} \pm \num{0.0480}$
        \\
        dsCEM   & $\num{0.1669} \pm \num{0.0454}$ & $\num{0.0634} \pm \num{0.0358}$
        \\
        \bottomrule
        \multicolumn{3}{l}{\footnotesize $^{\text{a}}$mean $\pm$ standard deviation in \si{\milli\second} evaluated over 100 runs} \\
    \end{tabular}
    }
\end{table}

Computation times per control step are reported in \cref{tab:computation_times}.
All runtimes were measured on a single Intel Xeon Platinum 8368 core.
As expected, heuristic variants are generally faster than trust region variants because trust region updates require an additional optimization step.
Across sampling methods, deterministic variants (dsTR and dsMPPI) are faster than low-discrepancy variants (Halton and Sobol) due to precomputed samples, while random sampling remains the fastest.
Further runtime reductions are possible through parallelization.

\subsection{MuJoCo Playground Validation}

As a supplementary validation beyond the classical benchmark tasks, we evaluate the method on WalkerRun and CheetahRun from MuJoCo Playground~\cite{zakkaDemonstratingMuJoCoPlayground2025}. 
The control discretization is $\SI{0.05}{\second}$, the simulation step is $\SI{0.01}{\second}$, and each run spans \num{250} simulation steps. 
We use $N=\num{100}$ samples, $\epsilon = \num{3}$, $H_{\min} = \num{-75}$, and a horizon length of $\horizon=10$. The initial proposal is a Gaussian $\Gaussian(\vzero, \mI)$, transformed with $\beta=1$ (see \cref{sec:standard_improvements}) to introduce temporal correlations; other controller settings match the previous experiments.
Both environments use six-dimensional control inputs, resulting in a 60-dimensional optimization vector over the horizon.
Each experiment is repeated for \num{100} randomized initial states.

\begin{figure}%
    \centering%
        \includegraphics[width=\linewidth]{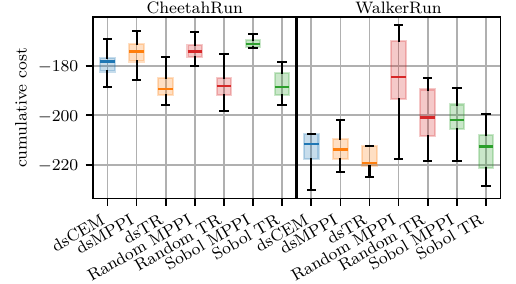}
    \caption{MuJoCo Playground validation results.}
    \label{fig:mujoco}%
\end{figure}%

The results in \cref{fig:mujoco} are consistent with the benchmark study: trust-region variants outperform the heuristic baselines, and the deterministic \ac{LCD}-based trust-region method achieves the lowest median cost, with the largest improvement in WalkerRun.

\section{Discussion}
\label{sec:discussion}

The experiments show that trust region updates improve both sample efficiency and convergence compared to heuristic updates across all considered settings and sampling methods.
Although heuristic methods remain competitive in certain settings, the proposed trust region \ac{MPC} is consistently outperforming them, in particular, when computational budgets are limited, \ie, when only a small number of samples and iterations are feasible.
The entropy lower bound further serves as a safeguard against premature proposal collapse in longer optimization runs, where the proposal can otherwise contract too quickly and lose exploration before reaching the optimum.

Note that reducing the required sample size directly decreases computation time and memory usage. 
On parallel hardware, even a difference of a single sample can significantly affect the runtime. 
\Eg, if the hardware can process up to 32 trajectories (samples) in parallel, requiring 33 samples would effectively double the computation time.

Regarding sampling methods, low-discrepancy sampling (Halton and Sobol) generally improves upon random sampling, and deterministic \ac{LCD} sampling yields further improvements, particularly when combined with trust region updates.
The supplementary MuJoCo Playground validation is consistent with this trend: trust-region variants remain competitive in higher-dimensional tasks, with the deterministic \ac{LCD}-based trust-region method achieving the lowest median cost in the evaluated settings.

Notably, trust regions in combination with \ac{LCD}-based sampling improve both sample and iteration efficiency without sacrificing control smoothness, making it particularly attractive for online \ac{MPC} on resource-constrained hardware.

\section{Conclusion}
\label{sec:conclusion}

We proposed a trust region formulation for sampling-based \ac{MPC} that uses principled updates of proposal distribution parameters, eliminating the need for heuristic adaptation, commonly found in state-of-the-art methods such as \ac{MPPI}.
Across the considered experiments, the proposed methods demonstrated improved convergence behavior and sample efficiency compared to heuristic baselines.
Additionally, combining trust region updates with deterministic \ac{LCD}-based sampling yielded the strongest results in the low-sample and low-iteration regimes, further improving performance.
These results suggest that the proposed approach is well-suited for real-world applications, where reliable convergence and sample efficiency are essential.
\appendix
\subsection{Derivation of the Optimal Next Proposal}
\label[appendix]{sec:appendix:lagrangian}

Using the \ac{KL} divergence and entropy, given by
\begin{align}
    D_{\mathrm{KL}}(\pdf_{j+1} \| \pdf_{j})
    &= \int_{\IR^{\dimVarCem}} \pdf_{j+1} (\vVarCem) \log \pdf_{j+1} (\vVarCem) \dd \vVarCem
    \\
    &\qquad
    - \int_{\IR^{\dimVarCem}} \pdf_{j+1} (\vVarCem) \log \pdf_{j} (\vVarCem) \dd \vVarCem \enspace, \\
    H(\pdf_{j+1})
    &= -\int_{\IR^{\dimVarCem}} \pdf_{j+1} (\vVarCem) \log \pdf_{j+1} (\vVarCem) \dd \vVarCem \enspace,
\end{align}
the expanded Lagrangian~\cref{eq:tr:lagrangian} reads
\begin{align}
    &\mathcal{L}(\pdf_{j+1},\eta,\alpha, \nu)
    = \int_{\IR^{\dimVarCem}} \pdf_{j+1}(\vVarCem) J(\vVarCem) \, \dd \vVarCem \\
    &\qquad + (\eta + \alpha) \int_{\IR^{\dimVarCem}} \pdf_{j+1}(\vVarCem) \log \pdf_{j+1}(\vVarCem) \dd \vVarCem \\
    &\qquad - \eta \int_{\IR^{\dimVarCem}} \pdf_{j+1}(\vVarCem) \log \pdf_{j}(\vVarCem) \dd \vVarCem \\
    &\qquad - \eta \epsilon + \alpha H_{\min} 
    + \nu\left(\int_{\IR^{\dimVarCem}} \pdf_{j+1}(\vVarCem) \dd  \vVarCem - 1\right) \enspace.
\end{align}
Taking the functional derivative point-wise w.r.t. $\pdf_{j+1}$ yields
\begin{align}
    \label{eq:lagrange_func_derivative}
    \frac{\delta \mathcal{L}}{\delta \pdf_{j+1}}
    & = J(\vVarCem) + (\eta + \alpha) \left( \log \pdf_{j+1}(\vVarCem) + 1 \right) 
    \\
    & \qquad
    - \eta \log \pdf_{j}(\vVarCem) + \nu \enspace.
\end{align}
Setting this derivative to zero and solving for $\log \pdf_{j+1}$ gives
\begin{align}
    \label{eq:log_opt_proposal}
    \log \pdf_{j+1}^{\ast}(\vVarCem)
    = -\frac{J(\vVarCem)}{\eta + \alpha}
    + \frac{\eta}{\eta + \alpha} \log \pdf_{j}(\vVarCem)
    + C \enspace,
\end{align}
where $C = - \log Z_{\eta, \alpha}$ 
absorbs all additive constants. Exponentiating yields the optimal proposal $\pdf_{j+1}^{\ast}$ \cref{eq:opt_proposal}, where the multiplier $\nu$ is implicitly accounted for by $Z_{\eta, \alpha}$ required to satisfy the normalization constraint.

\subsection{Derivation of the Dual Function}
\label[appendix]{sec:appendix:dual}

To derive the dual function $g(\eta, \alpha)$, we substitute the optimal proposal $\pdf_{j+1}^{\ast}$ back into the Lagrangian. 
First, we take the logarithm of the optimal proposal \cref{eq:log_opt_proposal} multiplied by $\eta + \alpha$ (which is strictly positive, by \cref{thm:sum_positive}) to match the coefficient of the entropy term in the Lagrangian.
Taking the expectation of $(\eta + \alpha) \log \pdf_{j+1}^{\ast}(\vVarCem)$ w.r.t. $\pdf_{j+1}^{\ast}$ yields
\begin{align}
    &(\eta + \alpha) \int_{\IR^{\dimVarCem}} \pdf_{j+1}^{\ast}(\vVarCem) \log \pdf_{j+1}^{\ast}(\vVarCem) \dd \vVarCem 
    = 
    \\
    & \quad
    \eta \int \pdf_{j+1}^{\ast}(\vVarCem) \log \pdf_j(\vVarCem) \dd \vVarCem 
    - \int_{\IR^{\dimVarCem}} \pdf_{j+1}^{\ast}(\vVarCem) J(\vVarCem) \dd \vVarCem \\
    &\quad - (\eta + \alpha) \log Z_{\eta,\alpha} \int \pdf_{j+1}^{\ast}(\vVarCem) \dd \vVarCem \enspace.
    \label{eq:appendix:expected_log}
\end{align}
Since $\pdf_{j+1}^{\ast}$ is a valid \ac{PDF}, it integrates to $1$.
Then, substituting \cref{eq:appendix:expected_log} back into the simplified Lagrangian reads
\begin{align}
    &\mathcal{L}(\pdf_{j+1}^{\ast},\eta,\alpha) 
    = \int_{\IR^{\dimVarCem}} \pdf_{j+1}^{\ast}(\vVarCem) J(\vVarCem) \dd \vVarCem \nonumber \\
    & \quad + \eta \int_{\IR^{\dimVarCem}} \pdf_{j+1}^{\ast}(\vVarCem) \log \pdf_j(\vVarCem) \dd \vVarCem 
    \\
    & \quad - \int_{\IR^{\dimVarCem}} \pdf_{j+1}^{\ast}(\vVarCem) J(\vVarCem) \dd \vVarCem - (\eta + \alpha) \log Z_{\eta,\alpha} \nonumber \\
    &\quad - \eta \int_{\IR^{\dimVarCem}} \pdf_{j+1}^{\ast}(\vVarCem) \log \pdf_j(\vVarCem) \dd \vVarCem \nonumber \\
    &\quad - \eta \epsilon + \alpha H_{\min} \enspace,
\end{align}
where the normalization multiplier $\nu$ has been absorbed in $Z_{\eta, \alpha}$.
Notice that the terms involving the expected cost 
and the cross-entropy $\eta \int_{\IR^{\dimVarCem}} \pdf_{j+1}^{\ast}(\vVarCem) \log \pdf_j(\vVarCem) \dd \vVarCem$ cancel out. 
The remaining terms yield the dual function \cref{eq:tr_dual}.

\subsection{Strict Positivity of the Lagrange Multiplier Sum}
\label[appendix]{sec:appendix:pos_multiplier_sum}
\begin{theorem}
    \label{thm:sum_positive}
    Let $J$ be non-constant, $H_{\min} > -\infty$, $f_j$ be non-degenerate, and $\epsilon < \infty$. Then $\eta^{\ast} + \alpha^{\ast} > 0$.
\end{theorem}
\begin{proof}
By \ac{KKT} dual feasibility, $\eta^{\ast} \geq 0$ and $\alpha^{\ast} \geq 0$, so $\eta^{\ast} + \alpha^{\ast} = 0$ requires $\eta^{\ast} = \alpha^{\ast} = 0$.
Under this assumption, the Lagrangian~\cref{eq:tr:lagrangian} reduces to the linear functional $\int_{\IR^{\dimVarCem}} \pdf_{j+1}(\vVarCem) J(\vVarCem) \dd \vVarCem + \nu \left( \int_{\IR^{\dimVarCem}} \pdf_{j+1}(\vVarCem) \dd \vVarCem - 1 \right)$,
and \cref{eq:lagrange_func_derivative} reduces to $\nicefrac{\delta \mathcal{L}}{\delta \pdf_{j+1}} = J(\vVarCem) + \nu$.
Solving for $\pdf_{j+1}$ over valid probability measures requires either:
i) $J(\vVarCem) = -\nu$ everywhere on the support of $\pdf_{j+1}$, which requires $J$ to be constant if $\pdf_{j+1}$ is continuous and strictly positive; or
ii) $\pdf_{j+1} = \delta(\vVarCem - \vVarCem^*)$ with $\vVarCem^* \in \argmin_{\vVarCem} J(\vVarCem)$, concentrating all mass onto the global minimum of $J$.
The first case directly contradicts the assumption that $J$ is non-constant. The second case yields $H(\delta) = -\infty < H_{\min}$ and $D_{\mathrm{KL}}(\delta \| \pdf_j) = \infty > \epsilon$, where either violation suffices to contradict primal feasibility and \ac{KKT} optimality.
Therefore, $\eta^{\ast} = \alpha^{\ast} = 0$ cannot hold, implying $\eta^{\ast} + \alpha^{\ast} > 0$. 
\end{proof}
Hence, the exponents in \cref{eq:opt_proposal} never degenerate, regardless of which of the two constraints is active.

\subsection{Strong Duality}
\label[appendix]{sec:appendix:strong_duality}
We show strong duality under the following two weak assumptions, which hold for common setups such as positive-definite quadratic costs and Gaussian proposals.

\begin{assumption}
    \label{assumption:duality:on_cost}
    We consider minimizing the expected cost over a class of probability distributions $\mathcal{F}$.
    We assume that $J(\vVarCem)\ge c > -\infty,~ \forall \vVarCem \in \operatorfont{dom} (\mathcal{F})$ is chosen so that its expectation is bounded below by $c$ across the domain $\operatorfont{dom}(\mathcal{F})$ of the considered distributions in $\mathcal{F}$.
    This holds trivially without restricting $\mathcal{F}$ whenever $J(\vVarCem) \ge c > -\infty,~\forall \vVarCem$.
\end{assumption}
\begin{assumption}
    \label{assumption:duality:on_cost_and_proposal}
    The cost function and initial proposal distribution $\pdf_j \in \mathcal{F}$ are chosen such that $\Eop_{\pdf_j}\{J(\vVarCem)\} < +\infty$ holds.
\end{assumption}

To establish strong duality for \cref{eq:tr_objective_function}, we apply Luenberger's duality theorem for infinite-dimensional convex programs~\cite[Sec.~8.6, Thm.~1]{luenbergerOptimizationVectorSpace1969}, which generalizes Slater's condition from finite-dimensional vector spaces to function spaces. Under this theorem, strong duality holds if 1) the cost function and inequality constraints are convex, 2) there exists a strictly feasible point in $\pdf_{j+1}\in\mathcal{F}$, that is, there is a point where all inequality constraints hold strictly (excluding equality), and 3) the primal infimum ($d^*=\inf_{\pdf_{j+1\in \mathcal{F}, G\le0}} \Eop_{\pdf_{j+1}}\{J(\vVarCem)\}$) over the feasible domain of valid probability densities $\mathcal{F}$ is finite, where $G$ summarizes all inequality constraints.

We verify these requirements for \cref{eq:tr_objective_function} over $\mathcal{F}$, and by restricting $\mathcal{F}$ to valid distributions, we implicitly address the equality constraint for normalized \acp{PDF}.
\begin{enumerate}
    \item 
    \textbf{Convex structure:} Both the expected cost $\Eop_{\pdf_{j+1}}\{J(\vVarCem)\}$ and the normalization constraint are linear in $\pdf_{j+1}$. The inequality constraints $D_{\mathrm{KL}}(\pdf_{j+1} \Vert{} \pdf_j) - \epsilon \le 0$ and $H_{\min} - H(\pdf_{j+1}) \le 0$ are strictly convex functionals of $\pdf_{j+1}$, yielding convex feasible sets. The primal problem is thus a convex program.
    \item
    \textbf{Strict feasibility:} Setting $\pdf_{j+1} = \pdf_j \in \mathcal{F}$ yields a strictly feasible point, as $D_{\mathrm{KL}}(\pdf_j \Vert{} \pdf_j) = 0 < \epsilon$ and $H(\pdf_j) > H_{\min}$, where the latter holds by construction since the entropy floor is chosen strictly below the entropy of the current proposal.
    \item 
    \textbf{Finite primal infimum:} The primal infimum $d^*$ is finite due to three factors: 
    (i) the feasible set is non-empty (guaranteed by $\pdf_{j+1} = \pdf_j$);
    (ii) $+\infty$ is excluded since $\pdf_j$ satisfies $\Eop_{\pdf_j}\{J(\vVarCem)\} < +\infty$ by \cref{assumption:duality:on_cost_and_proposal}; and
    (iii) the expectation is bounded from below by \cref{assumption:duality:on_cost}.
\end{enumerate}

Since all conditions are satisfied in our setting, Luenberger's theorem for strong duality holds. 

\section*{Acknowledgment}

We express our sincere gratitude to Gerhard Neumann for discussions on trust regions in \ac{RL}.

\balance





\bibliographystyle{bib/IEEEtran_Capitalize}
\bibliography{IEEEabrv,bib/literature,bib/this_paper}


\end{document}